\documentclass[twocolumn,showpacs,footinbib,10pt, prl]{revtex4-1} 
\usepackage{color}    
\usepackage[hmargin=0.69in, vmargin=0.75in] {geometry}         
\usepackage[breaklinks]{hyperref}
\usepackage{graphicx}
\usepackage{amssymb}
\usepackage{amsmath}
\usepackage{epstopdf}
\usepackage{rotating}
\usepackage{natbib}
\usepackage[section]{placeins}
\usepackage[caption=false, position=t,singlelinecheck=off]{subfig}

\newcommand {\be}{\begin{equation}}
\newcommand {\ee}{\end{equation}}

\newcommand{\expect}[1]{\langle#1\rangle}

\begin{document}

\title{Polariton waveguides  from a quantum dot chain in  a nanowire photonic crystal:\\ an architecture for quantum information science and waveguide QED}
\author{Gerasimos Angelatos}
\email{g.angelatos@queensu.ca}
\author{Stephen Hughes}
\affiliation{Department of Physics, Engineering Physics and Astronomy, Queen's University, Kingston, Ontario, Canada K7L 3N6}
\begin{abstract}
We introduce a   polariton-waveguide structure, comprised of a nanowire-based photonic crystal waveguide  with a quantum dot embedded in each  unit cell.  Using realistic designs and parameters, we derive and calculate the fundamental electromagnetic properties of these polariton waveguides, with an emphasis on the photon Green function and local optical density of states (LDOS). Both infinite and finite-size waveguides are considered, where the latter's properties are calculated using a Dyson equation approach without any approximations. We demonstrate dramatic   increases,  and rich fundamental control, of the LDOS  due to strong light-matter interactions in each unit cell through periodic quantum dot interactions.  Consequently, these structures allow the exploration of new regimes of waveguide quantum electrodynamics.  As an example application,  we consider the coupling of an external target quantum dot with a finite-sized polariton waveguide, and show that the single quantum dot strong coupling regime is  easily accessible, even for   modest  dipole strengths.  
\end{abstract}
\maketitle


The design of nanoscale solid state devices to control light-matter interactions is highly desirable, both for applications in quantum information science \cite{Yao2009} and to explore novel regimes of quantum electrodynamics (QED) \cite{Fink2008, Greentree2006}.  Photonic crystals (PCs) with embedded quantum dots (QDs) have had much success in this regard by modifying the LDOS \cite{John1990, Hennessy2007,Gao2013, Bose2012}, although they have been hindered by  fabrication issues including  surface roughness \cite{Hughes2005, Patterson2009, Fussell2008} and limited control of the QD properties \cite{BaHoang2012}.  
Arrays of nanowires (NWs) grown through a molecular beam epitaxy (MBE) technique \cite{Dubrovskii2009, Joyce2011} have been proposed as an alternative PC platform to help mitigate these issues \cite{Angelatos2014, Angelatos2015}, with the potential to contain identical  or very similar QDs inside each NW of a given radius~\cite{Makhonin2013, Diedenhofen2011}.  This opens the idea of NW PC waveguides where each waveguide channel NW contains an identical QD embedded in its center.  Coupled dipole chains in free space have interesting collective properties, and can act as subwavelength waveguides \cite{Citrin2004}; implementing such chains in PC waveguides could result in new physical behavior or improved performance.  Metamaterials, comprised of multiple elements that are engineered to have unique and useful properties, have been used to produce exotic waveguides with a dramatically different LDOS than simple dielectric structures, e.g., manifesting in  large spontaneous emission rate enhancements \cite{Li2009, Yao2009meta} for dipole emitters, even with   material losses. Plasmonic polariton waveguide structures \cite{Christ2003} and metal nanoparticle chains coupled to traditional waveguides \cite{Fevrier2012} have demonstrated chain-mediated coupling between plasmonic and photonic excitations leading to an anti-crossing in the band structure.  This polariton anti-crossing is also observed with quantum wells in 1D distributed Bragg reflector waveguides \cite{Yablonskii2001}. All these structures essentially lead to normal mode splitting.

\begin{figure}[b]
\subfloat[\vspace{-1pt}]{\label{scheme}\includegraphics[width=0.23\textwidth]{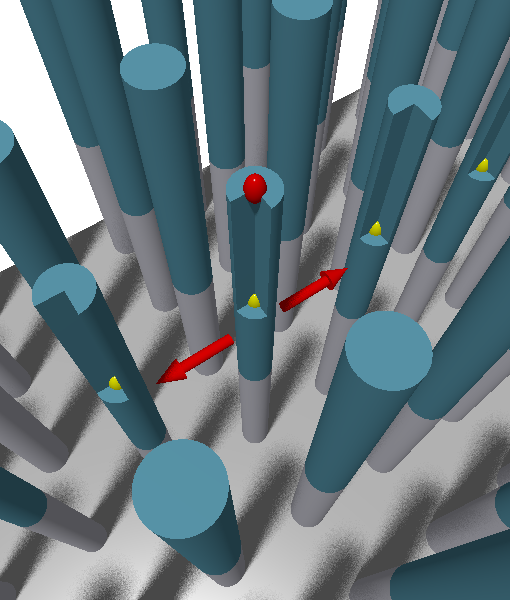}}\quad
\subfloat[\vspace{-1pt}]{\label{Gcomp}\includegraphics[width=0.23\textwidth]{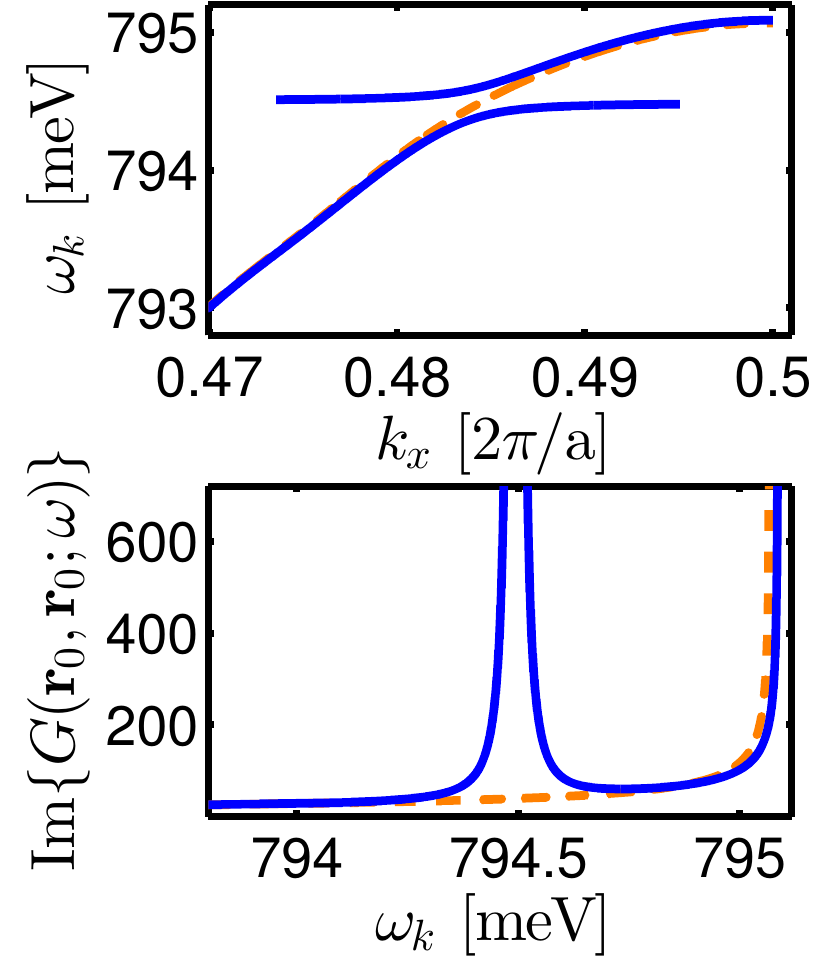}}
\vspace{-5pt}
\caption{(Color Online) (a)  Proposed polariton waveguide, with yellow QDs embedded inside the PC waveguide channel, and a red external QD coupling to the structure.  Red arrows denote the waveguide direction and NWs are cut out so embedded QDs can be seen.  (b) Slow light region of the band structure on top, and $\text{Im}\{G(\mathbf{r}_n, \mathbf{r}_n)\}$ in units of $\rho^h$ (see text) on bottom;   the infinite polariton waveguide in solid blue is compared with the original PC waveguide in dashed orange.
}
\label{inf}
\end{figure}

 Inspired by  the metamaterial concept and advances in  NW growth techniques, the inclusion of a QD chain in a PC waveguide can similarly reshape and greatly enhance the system LDOS and result in drastically different properties and behavior, but without the large metallic losses of plasmonic metamaterials.  In this Letter, we introduce and explore the optical properties of a nanophotonics system comprised of a PC waveguide with an embedded QD in each unit cell, which we describe as a ``polariton waveguide''  because its excitations are mixed light-matter states due to the collective coupling of the QDs with the waveguide Bloch mode. Figure \ref{inf}(a) shows a schematic of our proposed structure. To elucidate the underlying physics of these systems,  we derive the photon Green function (GF) of both infinite and finite-sized systems polariton waveguides,  and explore their coupling with a single external ``target'' QD, which is found to be in the strong coupling regime.  This achievement of strong coupling between a single isolated emitter and a 
\textit{waveguide mode}, to the best of our knowledge first reported here, 
allows for the creation of  devices relying on quantum cavity physics, such as photon blockades and single photon switches \cite{Greentree2006,  Bose2012}, with reliable input/output coupling \textit{on chip}.  Our polariton waveguide architecture can also be adapted for other systems such as  circuit QED \cite{Fink2008}.

Our proposed structure exploits  the elevated NW PC waveguide design of Ref.~\onlinecite{Angelatos2014}, where a PC waveguide is formed from an organized array of GaAs NWs extended from an AlO substrate \cite{Tokushima2004}, with the waveguide channel introduced by reducing the radius of a  row of NWs from $r_b=0.180\,a$ to $r_d=0.140\,a$ along $\mathbf{e}_x$. Importantly,  this structure is based on current fabrication techniques and properties are determined through full 3D calculations including radiative coupling effects.  The PC waveguide with lattice constant $a=0.5526\,\mu$m contains a single vertically-polarized below-light-line waveguide band with a mode edge near the  telecom wavelength of $1.550\,\mu$m (ranging from $755-795$\,meV).  Light is confined to the upper GaAs portion of the NWs (height $2.27\,a$), while the $2\,a$ AlO layer separates them from the substrate and an array width of $7\,a$ is sufficient to prevent in-plane losses.   As this waveguide band approaches the mode edge it flattens due to symmetry and the group velocity goes to zero, causing the LDOS to diverge as seen in Fig.~\ref{inf}\subref{Gcomp}, an effect that is inevitably spoiled by disorder-induced losses in real systems \cite{Hughes2005}.  To form the polariton waveguide of Fig.~\ref{inf}\subref{scheme}, we embed a QD in the center of the GaAs layer of each waveguide NW at $\mathbf{r}_{n, 0}=\mathbf{r}_0+na\,\mathbf{e}_x$, where $n$ is an integer.  We take the embedded QDs to have a Lorenzian polarizability  $\boldsymbol\alpha=\alpha(\omega)\mathbf{e}_z=2\omega_0|\mathbf{d}|^2\mathbf{e}_z/(\hbar\epsilon_0(\omega_0^2-\omega^2-i\Gamma_0\omega))$,  with a  dipole moment $|\mathbf{d}|=30$\,D  ($0.626\,$e-nm)   and polarization decay rate $\Gamma_0=1\,\mu$eV (including both non-radiative decay and coupling into non-waveguide modes); these are  similar to experimental parameters for InAs QDs at 4\,K~\cite{Weiler2012}. The QD exciton line is taken to be $\omega_0=794.5$\,meV, which is in the moderately-slow-light regime of the waveguide band resulting in a group velocity $v_g=c/30.4$ and  $F_z= 40.1$ (relative LDOS enhancement, $F_z=\text{Im}\{{G}\}/\text{Im}\{{G}^{\rm h}\}$). Throughout this work, we connect to the GF $\mathbf{G}(\mathbf{r}, \mathbf{r}'; \omega)$, which describes the system electric-field response at $\mathbf{r}$ to a point source at $\mathbf{r}'$ (fully including all light scattering events), and whose imaginary and real components at equal space points are directly proportional to the system LDOS  (and emitter spontaneous emission rate), and Lamb shift, respectively \cite{Yao2009, Angelatos2015}.  These GFs are projected along the relevant $\mathbf{e}_z$ component for vertically polarized QDs: $G=\mathbf{e}_z\cdot\mathbf{G}\cdot\mathbf{e}_z$ and given in units of the imaginary part of free-space GF  $\text{Im}\{{G}^{\rm h}(\mathbf{r}, \mathbf{r}; {\omega})\}=\omega^3/(6\pi c^3)\equiv \rho^h(\omega)$~\cite{Novotny2006}.  When we consider the addition of an external target QD to the waveguide, this will be embedded on the top of waveguide  NW at position $\mathbf{r}_n$, where $n$ indexes the unit cell, as seen in Fig.~\ref{inf}\subref{scheme}, both for fabrication purposes and because this is where the antinode of the waveguide mode resides.

We first consider an infinite embedded QD array to help explain these structures' underlying physics, exploiting Bloch's theorem and treating the QDs as a perturbation to each PC unit cell ${\Delta\boldsymbol\epsilon}(\mathbf{r})= \delta(\mathbf{r}-\mathbf{r}_0)\alpha(\omega_k')\mathbf{e}_z$, which shifts the waveguide resonance $\omega_k$ for a given $k\equiv k_x$  to $\omega_k'$.  We obtain from  perturbation theory, $\omega'^2_k= \omega_k^2(1-\int_{V_c}{\Delta\epsilon}(\mathbf{r})|\mathbf{e}_z\cdot\mathbf{u}_{{k}}(\mathbf{r})|^2 d\mathbf{r})$~\cite{Mahmoodian2009, Patterson2010, Angelatos2015th}, where $V_c$ is the unit cell volume and $\mathbf{u}_{{k}}(\mathbf{r})$ is the waveguide unit-cell function, normalized through $\int_{\rm V_c}\epsilon(\mathbf{r}) |\mathbf{u}_k(\mathbf{r})|^2d{\bf r }=1$,
 with corresponding waveguide mode $\mathbf{f}_k(\mathbf{r})=\sqrt{\frac{a}{L}}\mathbf{u}_{k}(\mathbf{r})e^{i k x}$ where $L$ the waveguide length.  
The waveguide band is split by the QD\textendash{}Bloch-mode interaction, resulting in a pair of complex eigenfrequencies ${\omega}'_{k, \pm}=\omega_{k,\pm} +i\Gamma_{k,\pm}/2$ at each $k$, with $\omega_{k,\pm}=\frac{\omega_0+\omega_k}{2}\pm\frac{1}{2}\sqrt{\left(\omega_0-\omega_k\right)^2+4g_k^2}$ and $\Gamma_{k, \pm}=\Gamma_0\frac{\omega_{k, \pm}^2-\omega_k^2}{\left(\omega_{k, \pm}^2-\omega_k^2\right)+\left(\omega_{k, \pm}^2-\omega_0^2\right)}$.   The QD\textendash{}unit-cell coupling parameter $g_k=
\sqrt{\frac{ \omega_0}{2\hbar\epsilon_0}}\mathbf{d}\cdot\mathbf{u}_{{k}}(\mathbf{r}_0)$ gives the strength of the light-matter interaction leading to anti-crossing and is the continuum form of $g$, the single mode quantum optical coupling rate.  This modified band structure is compared with that of the original PC in Fig.~\ref{inf}\subref{Gcomp}, where the strong mode splitting ($g_{k_{0}}=120\,\mu\text{eV}\gg\Gamma_0$ ) flattens the dispersion as $\omega_{k,\pm}\to\omega_0$ and losses remain low: $\Gamma_{k,\pm}\leq\Gamma_0\ll\omega_{k,\pm}$.  The inclusion of the QD array has caused the waveguide band to split, corresponding to a mixed light-matter excitation:  the {\it polariton waveguide}.

The above approach was also applied to the waveguide Bloch modes: away from  $\omega_{k, \pm}\to \omega_0$ where perturbation theory naturally breaks down we find only the waveguide Bloch mode contribution is significant and $\mathbf{u}_{ {k}, \pm}= \mathbf{u}_{ {k}}$, and we only show results for frequencies where perturbation theory still holds.
We write the GF of this infinite polariton waveguide as a sum over these waveguide Bloch modes, $\mathbf{G}_{\rm P}(\mathbf{r}, \mathbf{r}'; \omega)\!=\!\sum_{k, \pm} \frac{ {\omega'}_\pm(k)^2 \mathbf{f}_{k,\pm}(\mathbf{r}) \mathbf{f}^*_{k, \pm}(\mathbf{r}')}{{\omega'}_\pm(k)^2-\omega^2}$, and converting this sum to an integral in the complex plane~\cite{Angelatos2015th}, we  arrive at an analytic expression similar in form to that for a regular PC waveguide \cite{Rao2007theory, Yao2009}:
\begin{align}
\mathbf{G}_{\rm P}(\mathbf{r}, \mathbf{r'}; \omega)\!=& \!\frac{i a \omega}{2 \tilde{v}_g}'\Big[ \Theta(x-x')\mathbf{u}_{k_\omega}(\mathbf{r})\mathbf{u}^*_{ k_\omega}(\mathbf{r}')e^{(i k_\omega-\kappa_{\omega}) (x-x') } \nonumber \\
+\Theta&(x'-x)\mathbf{u}^*_{k_\omega}(\mathbf{r})\mathbf{u}_{ k_\omega}(\mathbf{r}')e^{ (ik_\omega-\kappa_{ \omega}) (x'-x) } \Big], \!\!\!\!
\label{eq:Gwguc}
\end{align}
but with $\tilde{v}_g'(\omega)=v'_g(\omega)-\frac{i}{2}\frac{d\Gamma_\pm(k)}{dk}|_{k_\omega}$, where $v'_g(\omega)\gg\frac{d\Gamma_\pm}{dk}$ is the polariton waveguide group velocity and $\kappa_{ \omega}=\frac{\Gamma_\pm(k_\omega)}{2v'_g(\omega)}$.  We emphasize that both $v'_g(\omega)$ and $k_\omega$ are found from the polariton waveguide $\omega_\pm-k$ relationship, which differs greatly from that of the original PC near $\omega_0$ due to the polariton splitting.  This was derived assuming $\kappa_\omega \ll k_{\omega}$, which remains valid until $v'_g\approx c/2300$, where $|\omega-\omega_0|<20\,\mu\text{eV}$.  The effect of the QD array, demonstrated in Fig.~\ref{inf}\subref{Gcomp}, is thus to produce large and tunable LDOS enhancements near resonance by flattening the band structure; as one moves away from $\omega_0,$ the original PC waveguide GF is quickly recovered.   At the above minimum detuning, these polariton waveguides yield $F_z=3100$---a dramatic rate enhancement and indicative that these structures should enter the strong coupling regime of QED, even at the single quantum level.

\begin{figure}
\subfloat[\vspace{-5pt}]{\label{incn2}\includegraphics[width=0.23\textwidth]{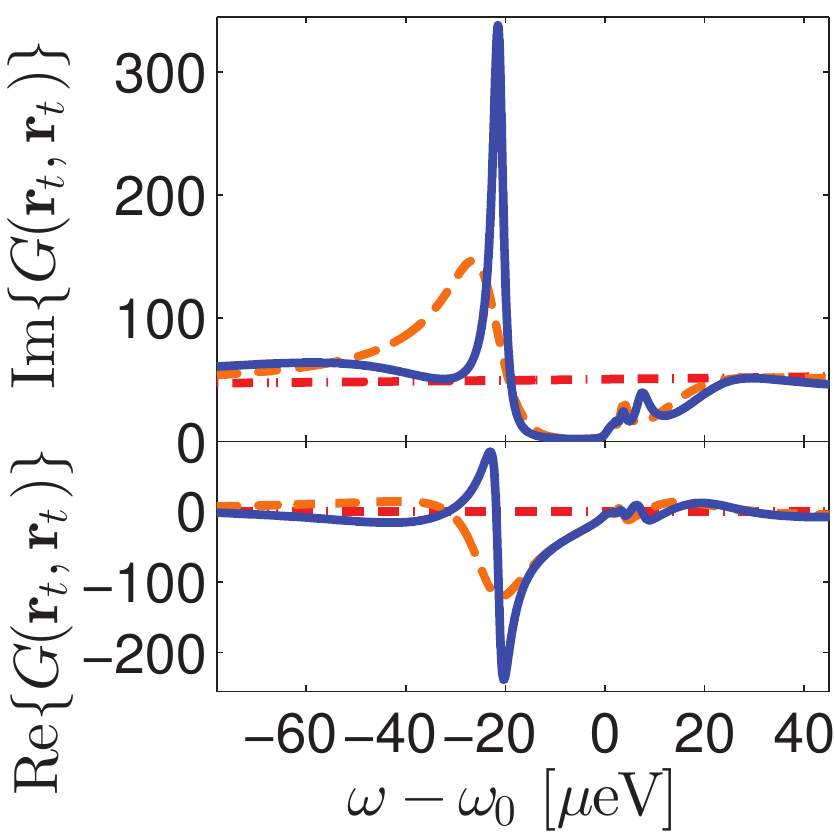}} \quad
\subfloat[\vspace{-1pt}]{\label{Gprop101}\includegraphics[width=0.23\textwidth]{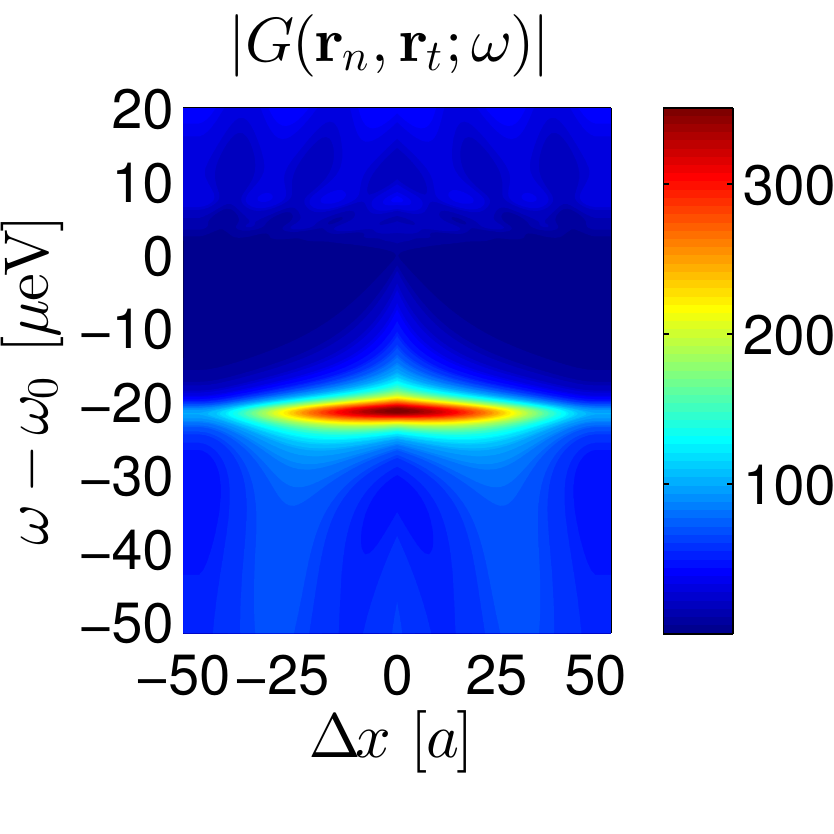}}
\vspace{-5pt}
\caption{(Color online) (a) Real and imaginary components of ${G}^{(101)}(\mathbf{r}_{51}, \mathbf{r}_{51}; \omega)$ in solid blue, compared with ${G}^{(51)}(\mathbf{r}_{26}, \mathbf{r}_{26}; \omega)$ in dashed orange and $N=0$ results in dash-dotted red.   (b) $|{G}^{(101)} ( \mathbf{r}_n, \mathbf{r}_{51}; \omega)|$ for $N=101$ polariton waveguide near  $\omega_0$, where $\Delta n=n-51$ and the array ends at $\Delta n=\pm50$. All values are in units of $\rho^h{(\omega)}$. }
\label{G101}
\end{figure}

We next consider polariton waveguides with  a finite number of embedded QDs, more representative  of real systems.  We can no longer exploit Bloch's theorem, but instead include QDs iteratively in the system GF via an exact Dyson equation approach.  Denoting the {\it background} PC waveguide Green function as $\mathbf{G}^{(0)}$ and starting with unit cell $n=1$, we introduce a QD at  $\mathbf{r}_{n, 0}$, calculate the resultant Green function  $\mathbf{G}^{(n)}$ including scattering from this QD, and then use this as the background Green function to introduce a QD in the subsequent unit cell.  From the Dyson equation, QD $n$ can be included self-consistently in the system GF through
 \cite{Kristensen2011}
$
\mathbf{G}^{(n)}(\mathbf{r}, \mathbf{r}')=\mathbf{G}^{(n-1)}(\mathbf{r}, \mathbf{r}')+ \mathbf{G}^{(n-1)}(\mathbf{r}, \mathbf{r}_{n, 0})\cdot
\boldsymbol\alpha \cdot\mathbf{G}^{(n)}(\mathbf{r}_{n, 0}, \mathbf{r}').
$
This was done numerically using the same system as for the infinite case (i.e.~identical PC waveguide and QDs) for QD chains of increasing length, $N=1\to101$.

Figure~\ref{G101} shows $G^{(101)}$, corresponding to a chain sufficiently long to produce substantial LDOS enhancements and begin to recover the infinite chain result, while still demonstrating important finite-size effects. 
We chose to emphasize the position on top of the central NW of the QD array $\mathbf{r}_{51}$ (where a target QD will be later embedded) because constructive interference from repeated QD scattering maximizes the polariton effects.  The addition of a single QD introduces a dip in $G$ at $\omega_0$, however with increasing $N$ the build up of off-resonant enhancement from QD scattering leads to the formation of a strong resonance in  ${\rm Im}\{G\}$ which exceeds $G^{(0)}$ for $N>15$. This resonance is red-shifted from $\omega_0$, with a peak $F_z=338.14$ and  FWHM $\Gamma_1=3.35\,\mu$eV at $\omega_1=\omega_0-21.49\,\mu$eV for the 101 QD case. As more QDs are added, this polariton peak grows, narrows, and blue-shifts towards $\omega_0$ while additional weaker red-shifted resonances begin to appear.  Above $\omega_0$, resonances also form which grow and  become more numerous with increasing $N$, indicating that they arise from QD chain Fabry-P\'{e}rot (FP) modes.  The higher two FP modes for the $N=101$ structure are at $\omega_{\rm FP'}=\omega_1+25.12\,\mu$eV and $\omega_{\rm FP}=\omega_1+29.02\,\mu$eV.  Away from $\omega_0$, $G^{(N)}$  converges to the PC waveguide GF, as was seen for the infinite case.  The large number of resonances in these waveguides also produce a richly varying ${\rm Re}\{G\}$ which remains substantial over a broad frequency region, particularly for larger $N$.

  The above results were also found to be robust to disorder in embedded QD position and dipole moment; e.g., random variations of up to $10\,$nm and 3\,D reduced the strength of the primary peak slightly ($\sim2\%$) and introduced a frequency shift of  $\sim1\,\mu$eV, but the underlying system behavior remained unchanged.  Similarly, these findings were also verified to hold for $\Gamma_0=0-10\,\mu$eV, with the $\omega_0$ dip extending deeper and resonances initially appearing at lower $N$ and closer to $\omega_0$ as $\Gamma_0$ is reduced.  Figure \ref{G101}\subref{Gprop101} shows the propagator $|{G}^{(101)} ( \mathbf{r}_n, \mathbf{r}_{51}; \omega)|$, which is proportional to the coupling strength between the target QD location and various points in the structure, and far greater coupling rates are found than for  a bare ideal PC waveguide.  For instance, $|{\rm Im}\{G^{(101)}(\mathbf{r}_n, \mathbf{r}_{51}; \omega)\}|>300\,\rho^h(\omega)$ and  $|{\rm Re}\{G^{(101)}(\mathbf{r}_n, \mathbf{r}_{51}; \omega)\}|>150\,\rho^h(\omega)$ are found for $\Delta n\leq 15$, and one can produce effectively any arbitrary combination of ${\rm Im}\{G\}$ and ${\rm Re}\{G\}$ through careful choice of separation and frequency.  This GF reshaping persists even past the mode edge of the polariton waveguide: once one is outside of the QD chain, $\mathbf{G}^{(N)}(\mathbf{r}_n, \mathbf{r}_{n'}; \omega)$ experiences only a phase shift of $e^{i k_\omega a|n-n'|}$ as $n$ and $n'$ are varied and $|G^{(101)}(\mathbf{r}_n, \mathbf{r}_{51}; \omega)|$ has a peak of $101.8\,\rho^h(\omega)$ at $\approx\omega_1$ ($\omega_0-21.6\,\mu$eV), more than double the bare PC waveguide result: $|G^{(0)}(\mathbf{r}_n, \mathbf{r}_{n'}; \omega_1)|=49.5\,\rho^h(\omega_1)$.   Notably these finite-sized GFs are derived without any approximations, and reproduce the physics of the infinite structure, namely strong light-matter interactions resulting in dramatic enhancements in the system LDOS.  

We now consider an important  quantum optical application of these polariton waveguides, studying the interaction of a single external QD at $\mathbf{r}_t=\mathbf{r}_{51}$ of the $N=101$ polariton waveguide.   The LDOS enhancements, specifically the polariton peak at $\omega_1$, are possibly sufficient to strongly couple the waveguide to a realistic QD.  We follow a first principals quantization procedure appropriate for lossy inhomogeneous systems \cite{Suttorp2004} such as these polariton waveguides.  The system Hamiltonian in the dipole approximation, consisting of a single target QD interacting with the electromagnetic environment of the polariton waveguide, is given by 
${H}=\hbar\omega_t \hat{\sigma}^+\hat{\sigma}^-+ \int  d\mathbf{r} \int_0^\infty d\omega_l\,\hbar \omega_l \hat{\mathbf{b}}^\dagger (\mathbf{r}; \omega_l)\cdot
\hat{\mathbf{b}}(\mathbf{r}; \omega_l) 
-(\hat{\sigma}^++\hat{\sigma}^-)\big( \mathbf{d}_t
 \cdot\hat{\mathbf{E}}(\mathbf{r}_t) + {\rm H.c.}\big),$ where the QD is a two-level atom (TLA) with exciton frequency $\omega_t$ and transition dipole moment $\mathbf{d}_t$;  $\hat{\mathbf{b}}(\mathbf{r}; \omega_l)$ is the  bosonic field annihilation operator associated with the field frequency $\omega_l$ which generates the electric-field operator via $\hat{\mathbf{E}}(\mathbf{r})
 \!\propto\! 
\int d\omega_l\int d\mathbf{r}'\sqrt{\text{Im}\{\epsilon(\mathbf{r}'; \omega_l) \}}\mathbf{G}(\mathbf{r}, \mathbf{r}'; \omega_l)\cdot\hat{\mathbf{b}}(\mathbf{r}'; \omega_l)$ \cite{Suttorp2004}.  We Laplace transform  the resulting operator Heisenberg equations of motion  to construct the spontaneous emission spectrum  at detector position $\mathbf{r}_D$:
$S(\mathbf{r}_D, \omega)= \expect{\hat{\mathbf{E}}^\dagger(\mathbf{r}_D, \omega)\hat{\mathbf{E}}(\mathbf{r}_D, \omega)}$, which yields
 \be
S(\mathbf{r}_D, \omega)=
\left|\frac{(\omega_t+\omega)\mathbf{G} ( \mathbf{r}_D, \mathbf{r}_t; \omega)\cdot\mathbf{d}/\epsilon_0}
{\omega_t^2-\omega^2-\omega\Sigma(\omega)-i\omega\Gamma_t}\right|^2,
\label{eq:spontespec}
\ee
 where we have assumed an initially excited QD in vacuum.
The self-energy is $\Sigma(\omega)=\frac{2\mathbf{d}_t\cdot\mathbf{G} ( \mathbf{r}_t, \mathbf{r}_t; \omega)\cdot\mathbf{d}_t}{\hbar\epsilon_0}$, and $\Gamma_t$ is the polarization decay rate of the target TLA due to interactions with the environment  (i.e., anything other than the LDOS given by the polariton waveguide GF).  
The spectrum depends not only on the GF at TLA position $\mathbf{r}_t$ but also on the propagator to the detector $\mathbf{G} ( \mathbf{r}_D, \mathbf{r}_t; \omega)$.  When we present spectra below, we show both $S(\mathbf{r}_D, \omega)$ and $S^0(\omega)$, where $S^0( \omega)$ is calculated from Eq.~\eqref{eq:spontespec} by replacing the propagator in the numerator with $\mathbf{G} ( \mathbf{r}_t, \mathbf{r}_t; \omega_t)$.  This is done because $S(\mathbf{r}_D, \omega)$ is often distorted by the rich spectral features of the GFs of these waveguide structures (see Fig.~\ref{Gprop101}).  The bare spectrum $S^0( \omega)$ thus more cleanly shows the system energy levels and dynamics of the  target QD.

\begin{figure}[t]
\centering
\vspace{-15pt}
\subfloat[\vspace{-1pt}]{\label{scspecres}\includegraphics[width=0.23\textwidth]{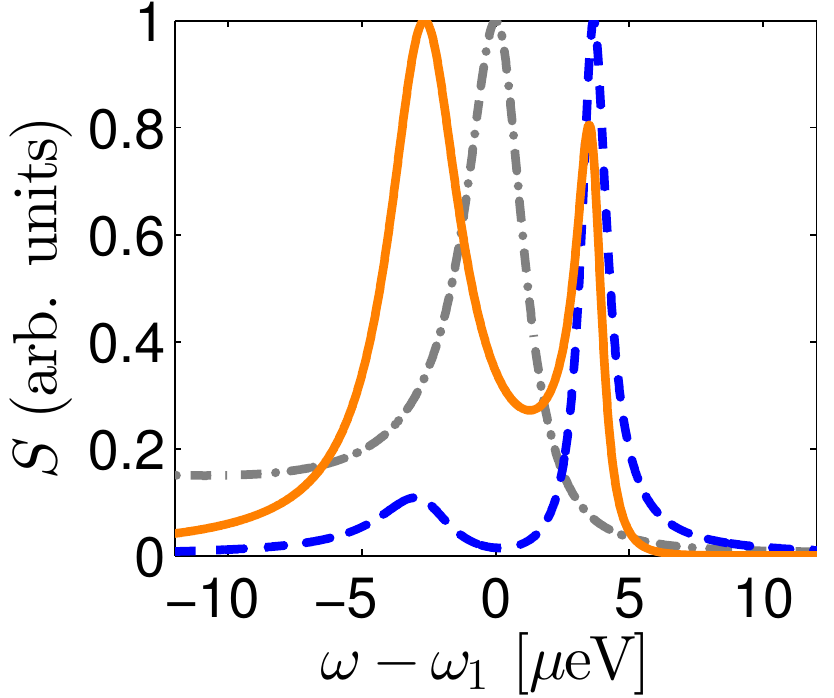}}\quad
\subfloat[\vspace{-1pt}]{\label{scspecfpb}\includegraphics[width=0.23\textwidth]{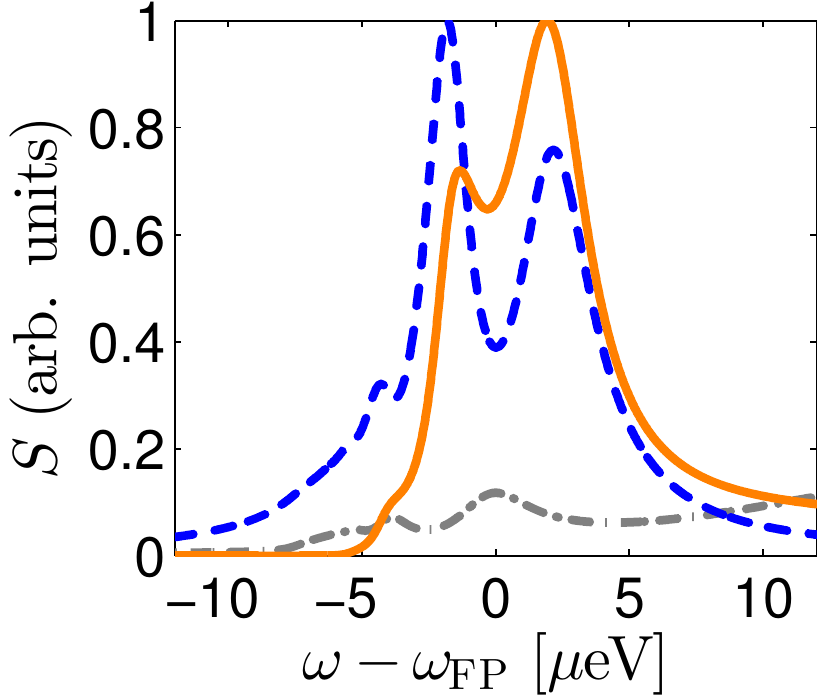}}
\vspace{-5pt}
\caption
{(Color online) $S^0(\omega)$   in dashed blue and $S(\mathbf{r}_D, \omega)$ in solid orange  for $\omega_t=\omega_{1}$, with  $d_t=30$\,D in (a) and $\omega_t=\omega_{\rm FP}$ with $d_t=60$\,D in (b). $\text{Im}\{G(\mathbf{r}_t, \mathbf{r}_t; \omega)\}$ in arbitrary units is  shown in dash-dotted gray.}
\label{specs}
\end{figure}

We assume a $\mathbf{e}_z$-aligned target QD with a polarization decay rate $\Gamma_t=1\,\mu$eV and calculate spectra for target QD dipole moments of $10$, $30$, and $60\,$D, encompassing the range of QDs which could feasibly be coupled to the polariton waveguide.  For a simple Lorentzian LDOS, the resultant system dynamics can be approximated with  the Jaynes-Cummings (JC) Hamiltonian, where the interaction of a single photonic mode and  TLA leads to an anti-crossing as they approach resonance (splitting of $2\hbar g$ on resonance); near $\omega_1$ we define an effective coupling constant $|g_{\rm eff}|=\sqrt{\frac{\Gamma_1\mathbf{d}_t\cdot \text{Im}\{\mathbf{G}(\mathbf{r}_t, \mathbf{r}_t; \omega_1)\}\cdot\mathbf{d}_t}{2\hbar\epsilon_0}}$ \cite{Angelatos2015th}. The three target dipole moments produce $g_{\rm eff}$ of  1.24, 3.72, and $7.43\,\mu$eV, respectively, which meet the criteria for strongly coupling with the $\omega_1$ resonance: $2g_{\rm eff}>\Gamma_t, \Gamma_1$ \cite{Meystre1999}. We assign a detector position $\mathbf{r}_D=\mathbf{r}_t-55\,a\,\mathbf{e}_x$, which is outside the polariton waveguide portion of the structure to reduce the filtering of the detected spectrum, while in the same position in the unit cell as the QD to maximize coupling.  The emitted and detected spectra $S^0(\omega)$ and $S(\mathbf{r}_D, \omega)$ at $\omega_t=\omega_1$ for the $30\,$D QD is shown in Fig.~\ref{specs}\subref{scspecres}. Remarkably, substantial splitting is seen  in both  $S^0(\omega)$ and $S(\mathbf{r}_D, \omega)$ despite the modest choice of $d_t$ and peak locations of $\omega_-=\omega_1-3.02\,\mu$eV and $\omega_+=\omega_1+3.67\,\mu$eV agree well with the JC Hamiltonian when the Lamb shift is included; however we stress that the widths and weighting of the spectral peaks, as well as propagation effects and polarization decay, are only captured through the formalism leading to Eq.~\ref{eq:spontespec}; in particular, the reduced height of the $\omega_-$ peak is a result of the unique shape of $\text{Re}\{G\}$ near $\omega_1$ producing a large positive Lamb shift.  The importance of propagation effects can clearly be seen by comparing  $S^0(\omega)$ and $S(\mathbf{r}_D, \omega)$, where the depletion above $\omega_1$ reduces the height of the $\omega_+$ peak.  In Fig.~\ref{specs}\subref{scspecfpb} we show the spontaneous emission spectra at $\omega_t=\omega_{\rm FP}$ for $d_t=60\,$D, where the QD\textendash{}polariton-waveguide coupling is sufficiently strong that significant exchange occurs with both the $\omega_{\rm FP}$ and $\omega_{\rm FP'}$ modes for $\omega_t=\omega_{\rm FP}$.  This  results  in  a triplet forming in $S^0(\omega)$, with peaks at $\omega_{\rm FP'}-0.44\,\mu$eV,  $\omega_{\rm FP}-1.81\,\mu$eV, and $\omega_{\rm FP}+2.16\,\mu$eV; the splitting is substantially stronger than  $\Gamma_{\rm FP}/2$ so these peaks are clearly seen at the detector position as well.

\begin{figure}[t]
\centering
\vspace{-5pt}
\includegraphics[width=0.48\textwidth]{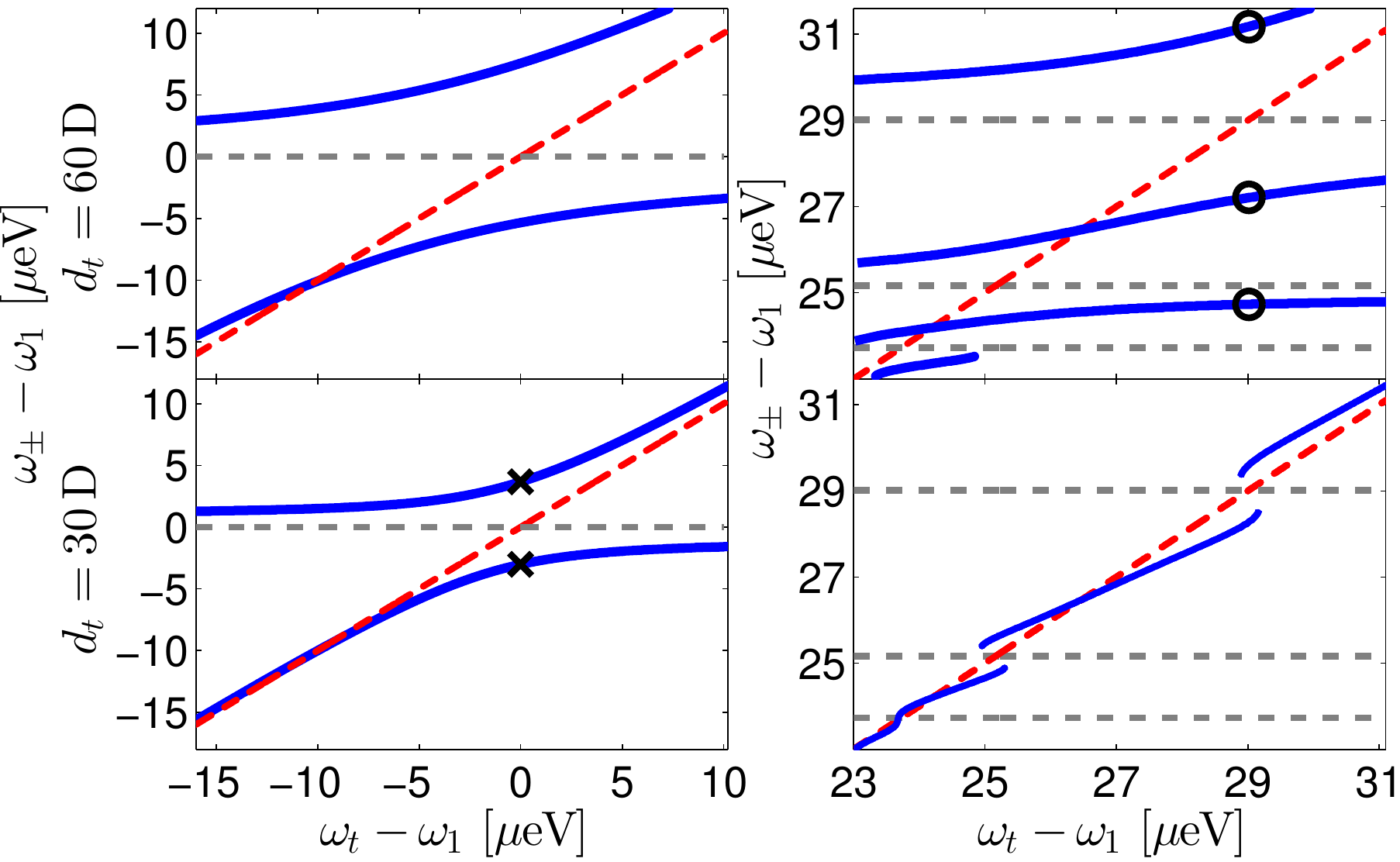}
\vspace{-5pt}
\caption{(Color online) Peaks in $S^0(\omega)$ in blue as a function of $\omega_t$ and  $d_t$.  Results for $d_t=30$ and $60\,$D are presented on the bottom and top, with $\omega_t$ near $\omega_1$ and in the FP region on the left and right, respectively.   Crosses and circles denote peaks of Figs.~\protect\ref{specs}\protect\subref{scspecres} and \protect\subref{scspecfpb},  $\omega_t$ is in dashed red, and LDOS peaks are in dashed gray.}
\label{scanticross}
\end{figure}

Finally, to demonstrate that the features in the above spontaneous emission spectra  are a consequence of the strong coupling regime, the clearly resolvable peaks of $S^0(\omega)$ as  $\omega_t$ is brought to resonance with $\omega_1$ are shown in Fig.~\ref{scanticross} for a target QD with $d_t=30$ and $60\,$D.  The left plots depict anti-crossing with  the primary resonance at $\omega_1$, while the peaks  as $\omega_t$ is swept through the FP region are shown on the right and markers denote the peaks of Fig.~\ref{specs}.  Remarkably for a waveguide system, all three QDs ($d_t=10$\,D is not shown) demonstrate a clear anti-crossing as they approach $\omega_1$, conclusive evidence that they are strongly coupled to the polariton-waveguide despite the complicated nature of this system. While similar strong coupling behavior has been theoretically predicted  for coupled-cavity PC waveguides \cite{Fussell2007}, a  later study showed this was inevitably spoiled by disorder  \cite{Fussell2008}.  Strong coupling has recently been observed with Anderson-localized cavities in disordered PC waveguides \cite{Gao2013},  although this is fundamentally different from strong coupling with a propagating waveguide mode as reported here.  It is also impressive that our predicted splitting is seen even for the $10\,$D target QD, which is much weaker than that used in other studies.  Of further interest is the system behavior in the FP region of the polarition waveguide,  where the unusual shape of the system LDOS leads to multiple anti-crossings which are poorly described by a JC Hamiltonian.  The interaction of the FP modes with the $60\,$D target QD is particularly striking: up to four energy levels are seen for a given $\omega_t$ and the  multi-mode nature of the QD-polariton waveguide has flattened these anti-crossing lines.

In conclusion,  we have proposed a new nano-engineered metamaterial system, a polaritonic waveguide, consisting of a PC waveguide with periodic embedded QDs.  We developed two separate approaches to describe the physics of this system, studying both infinite and finite polariton waveguides and demonstrating that in both instances strong light-matter interactions lead to rich and dramatic LDOS enhancements which are robust to disorder and without the losses typically associated with metallic metamaterials.  We then considered the interaction of a finite-size structure with an external QD and we showed that these LDOS enhancements can be exploited to strongly couple with a single external emitter and produce interesting spectral features clearly visible in the externally detected  spectrum.  These  structures could thus  be used to design complex devices for quantum information and explore  new regimes of open system waveguide QED. 

This work was supported by the Natural Sciences and Engineering Research Council of Canada and Queen's University.\\\

\bibliography{p3bib}

\begin{thebibliography}{34}%
\makeatletter
\providecommand \@ifxundefined [1]{%
 \@ifx{#1\undefined}
}%
\providecommand \@ifnum [1]{%
 \ifnum #1\expandafter \@firstoftwo
 \else \expandafter \@secondoftwo
 \fi
}%
\providecommand \@ifx [1]{%
 \ifx #1\expandafter \@firstoftwo
 \else \expandafter \@secondoftwo
 \fi
}%
\providecommand \natexlab [1]{#1}%
\providecommand \enquote  [1]{``#1''}%
\providecommand \bibnamefont  [1]{#1}%
\providecommand \bibfnamefont [1]{#1}%
\providecommand \citenamefont [1]{#1}%
\providecommand \href@noop [0]{\@secondoftwo}%
\providecommand \href [0]{\begingroup \@sanitize@url \@href}%
\providecommand \@href[1]{\@@startlink{#1}\@@href}%
\providecommand \@@href[1]{\endgroup#1\@@endlink}%
\providecommand \@sanitize@url [0]{\catcode `\\12\catcode `\$12\catcode
  `\&12\catcode `\#12\catcode `\^12\catcode `\_12\catcode `\%12\relax}%
\providecommand \@@startlink[1]{}%
\providecommand \@@endlink[0]{}%
\providecommand \url  [0]{\begingroup\@sanitize@url \@url }%
\providecommand \@url [1]{\endgroup\@href {#1}{\urlprefix }}%
\providecommand \urlprefix  [0]{URL }%
\providecommand \Eprint [0]{\href }%
\providecommand \doibase [0]{http://dx.doi.org/}%
\providecommand \selectlanguage [0]{\@gobble}%
\providecommand \bibinfo  [0]{\@secondoftwo}%
\providecommand \bibfield  [0]{\@secondoftwo}%
\providecommand \translation [1]{[#1]}%
\providecommand \BibitemOpen [0]{}%
\providecommand \bibitemStop [0]{}%
\providecommand \bibitemNoStop [0]{.\EOS\space}%
\providecommand \EOS [0]{\spacefactor3000\relax}%
\providecommand \BibitemShut  [1]{\csname bibitem#1\endcsname}%
\let\auto@bib@innerbib\@empty
\bibitem [{\citenamefont {Yao}\ \emph {et~al.}(2009{\natexlab{a}})\citenamefont
  {Yao}, \citenamefont {{Manga Rao}},\ and\ \citenamefont {Hughes}}]{Yao2009}%
  \BibitemOpen
  \bibfield  {author} {\bibinfo {author} {\bibfnamefont {P.}~\bibnamefont
  {Yao}}, \bibinfo {author} {\bibfnamefont {V.~S.~C.}\ \bibnamefont {{Manga
  Rao}}}, \ and\ \bibinfo {author} {\bibfnamefont {S.}~\bibnamefont {Hughes}},\
  }\href {\doibase 10.1002/lpor.200810081} {\bibfield  {journal} {\bibinfo
  {journal} {Laser Photon. Rev.}\ }\textbf {\bibinfo {volume} {4}},\ \bibinfo
  {pages} {499} (\bibinfo {year} {2009}{\natexlab{a}})}\BibitemShut {NoStop}%
\bibitem [{\citenamefont {Fink}\ \emph {et~al.}(2008)\citenamefont {Fink},
  \citenamefont {G\"{o}ppl}, \citenamefont {Baur}, \citenamefont {Bianchetti},
  \citenamefont {Leek}, \citenamefont {Blais},\ and\ \citenamefont
  {Wallraff}}]{Fink2008}%
  \BibitemOpen
  \bibfield  {author} {\bibinfo {author} {\bibfnamefont {J.~M.}\ \bibnamefont
  {Fink}}, \bibinfo {author} {\bibfnamefont {M.}~\bibnamefont {G\"{o}ppl}},
  \bibinfo {author} {\bibfnamefont {M.}~\bibnamefont {Baur}}, \bibinfo {author}
  {\bibfnamefont {R.}~\bibnamefont {Bianchetti}}, \bibinfo {author}
  {\bibfnamefont {P.~J.}\ \bibnamefont {Leek}}, \bibinfo {author} {\bibnamefont
  {Blais}}, \ and\ \bibinfo {author} {\bibfnamefont {A.}~\bibnamefont
  {Wallraff}},\ }\href {\doibase 10.1038/nature07112} {\bibfield  {journal}
  {\bibinfo  {journal} {Nature}\ }\textbf {\bibinfo {volume} {454}},\ \bibinfo
  {pages} {315} (\bibinfo {year} {2008})},\ \Eprint
  {http://arxiv.org/abs/0902.1827} {arXiv:0902.1827} \BibitemShut {NoStop}%
\bibitem [{\citenamefont {Greentree}\ \emph {et~al.}(2006)\citenamefont
  {Greentree}, \citenamefont {Tahan}, \citenamefont {Cole},\ and\ \citenamefont
  {Hollenberg}}]{Greentree2006}%
  \BibitemOpen
  \bibfield  {author} {\bibinfo {author} {\bibfnamefont {A.~D.}\ \bibnamefont
  {Greentree}}, \bibinfo {author} {\bibfnamefont {C.}~\bibnamefont {Tahan}},
  \bibinfo {author} {\bibfnamefont {J.~H.}\ \bibnamefont {Cole}}, \ and\
  \bibinfo {author} {\bibfnamefont {L.~C.~L.}\ \bibnamefont {Hollenberg}},\
  }\href {\doibase 10.1038/nphys466} {\bibfield  {journal} {\bibinfo  {journal}
  {Nat. Phys.}\ }\textbf {\bibinfo {volume} {2}},\ \bibinfo {pages} {856}
  (\bibinfo {year} {2006})}\BibitemShut {NoStop}%
\bibitem [{\citenamefont {John}\ and\ \citenamefont {Wang}(1990)}]{John1990}%
  \BibitemOpen
  \bibfield  {author} {\bibinfo {author} {\bibfnamefont {S.}~\bibnamefont
  {John}}\ and\ \bibinfo {author} {\bibfnamefont {J.}~\bibnamefont {Wang}},\
  }\href {\doibase 10.1103/PhysRevLett.64.2418} {\bibfield  {journal} {\bibinfo
   {journal} {Physical Review Letters}\ }\textbf {\bibinfo {volume} {64}},\
  \bibinfo {pages} {2418} (\bibinfo {year} {1990})}\BibitemShut {NoStop}%
\bibitem [{\citenamefont {Hennessy}\ \emph {et~al.}(2007)\citenamefont
  {Hennessy}, \citenamefont {Badolato}, \citenamefont {Winger}, \citenamefont
  {Gerace}, \citenamefont {Atat\"{u}re}, \citenamefont {Gulde}, \citenamefont
  {F\"{a}lt}, \citenamefont {Hu},\ and\ \citenamefont
  {Imamo\u{g}lu}}]{Hennessy2007}%
  \BibitemOpen
  \bibfield  {author} {\bibinfo {author} {\bibfnamefont {K.}~\bibnamefont
  {Hennessy}}, \bibinfo {author} {\bibfnamefont {A.}~\bibnamefont {Badolato}},
  \bibinfo {author} {\bibfnamefont {M.}~\bibnamefont {Winger}}, \bibinfo
  {author} {\bibfnamefont {D.}~\bibnamefont {Gerace}}, \bibinfo {author}
  {\bibfnamefont {M.}~\bibnamefont {Atat\"{u}re}}, \bibinfo {author}
  {\bibfnamefont {S.}~\bibnamefont {Gulde}}, \bibinfo {author} {\bibfnamefont
  {S.}~\bibnamefont {F\"{a}lt}}, \bibinfo {author} {\bibfnamefont {E.~L.}\
  \bibnamefont {Hu}}, \ and\ \bibinfo {author} {\bibfnamefont {A.}~\bibnamefont
  {Imamo\u{g}lu}},\ }\href {\doibase 10.1038/nature05586} {\bibfield  {journal}
  {\bibinfo  {journal} {Nature}\ }\textbf {\bibinfo {volume} {445}},\ \bibinfo
  {pages} {896} (\bibinfo {year} {2007})}\BibitemShut {NoStop}%
\bibitem [{\citenamefont {Gao}\ \emph {et~al.}(2013)\citenamefont {Gao},
  \citenamefont {Combrie}, \citenamefont {Liang}, \citenamefont
  {Schmitteckert}, \citenamefont {Lehoucq}, \citenamefont {Xavier},
  \citenamefont {Xu}, \citenamefont {Busch}, \citenamefont {Huffaker},
  \citenamefont {{De Rossi}},\ and\ \citenamefont {Wong}}]{Gao2013}%
  \BibitemOpen
  \bibfield  {author} {\bibinfo {author} {\bibfnamefont {J.}~\bibnamefont
  {Gao}}, \bibinfo {author} {\bibfnamefont {S.}~\bibnamefont {Combrie}},
  \bibinfo {author} {\bibfnamefont {B.}~\bibnamefont {Liang}}, \bibinfo
  {author} {\bibfnamefont {P.}~\bibnamefont {Schmitteckert}}, \bibinfo {author}
  {\bibfnamefont {G.}~\bibnamefont {Lehoucq}}, \bibinfo {author} {\bibfnamefont
  {S.}~\bibnamefont {Xavier}}, \bibinfo {author} {\bibfnamefont
  {X.}~\bibnamefont {Xu}}, \bibinfo {author} {\bibfnamefont {K.}~\bibnamefont
  {Busch}}, \bibinfo {author} {\bibfnamefont {D.~L.}\ \bibnamefont {Huffaker}},
  \bibinfo {author} {\bibfnamefont {A.}~\bibnamefont {{De Rossi}}}, \ and\
  \bibinfo {author} {\bibfnamefont {C.~W.}\ \bibnamefont {Wong}},\ }\href
  {\doibase 10.1038/srep01994} {\bibfield  {journal} {\bibinfo  {journal} {Sci.
  Rep.}\ }\textbf {\bibinfo {volume} {3}},\ \bibinfo {pages} {1994} (\bibinfo
  {year} {2013})}\BibitemShut {NoStop}%
\bibitem [{\citenamefont {Bose}\ \emph {et~al.}(2012)\citenamefont {Bose},
  \citenamefont {Sridharan}, \citenamefont {Kim}, \citenamefont {Solomon},\
  and\ \citenamefont {Waks}}]{Bose2012}%
  \BibitemOpen
  \bibfield  {author} {\bibinfo {author} {\bibfnamefont {R.}~\bibnamefont
  {Bose}}, \bibinfo {author} {\bibfnamefont {D.}~\bibnamefont {Sridharan}},
  \bibinfo {author} {\bibfnamefont {H.}~\bibnamefont {Kim}}, \bibinfo {author}
  {\bibfnamefont {G.~S.}\ \bibnamefont {Solomon}}, \ and\ \bibinfo {author}
  {\bibfnamefont {E.}~\bibnamefont {Waks}},\ }\href {\doibase
  10.1103/PhysRevLett.108.227402} {\bibfield  {journal} {\bibinfo  {journal}
  {Phys. Rev. Lett.}\ }\textbf {\bibinfo {volume} {108}},\ \bibinfo {pages}
  {227402} (\bibinfo {year} {2012})}\BibitemShut {NoStop}%
\bibitem [{\citenamefont {Hughes}\ \emph {et~al.}(2005)\citenamefont {Hughes},
  \citenamefont {Ramunno}, \citenamefont {Young},\ and\ \citenamefont
  {Sipe}}]{Hughes2005}%
  \BibitemOpen
  \bibfield  {author} {\bibinfo {author} {\bibfnamefont {S.}~\bibnamefont
  {Hughes}}, \bibinfo {author} {\bibfnamefont {L.}~\bibnamefont {Ramunno}},
  \bibinfo {author} {\bibfnamefont {J.~F.}\ \bibnamefont {Young}}, \ and\
  \bibinfo {author} {\bibfnamefont {J.~E.}\ \bibnamefont {Sipe}},\ }\href
  {\doibase 10.1103/PhysRevLett.94.033903} {\bibfield  {journal} {\bibinfo
  {journal} {Phys. Rev. Lett.}\ }\textbf {\bibinfo {volume} {94}},\ \bibinfo
  {pages} {033903} (\bibinfo {year} {2005})}\BibitemShut {NoStop}%
\bibitem [{\citenamefont {Patterson}\ \emph {et~al.}(2009)\citenamefont
  {Patterson}, \citenamefont {Hughes}, \citenamefont {Combri\'{e}},
  \citenamefont {Tran}, \citenamefont {{De Rossi}}, \citenamefont {Gabet},\
  and\ \citenamefont {Jaou\"{e}n}}]{Patterson2009}%
  \BibitemOpen
  \bibfield  {author} {\bibinfo {author} {\bibfnamefont {M.}~\bibnamefont
  {Patterson}}, \bibinfo {author} {\bibfnamefont {S.}~\bibnamefont {Hughes}},
  \bibinfo {author} {\bibfnamefont {S.}~\bibnamefont {Combri\'{e}}}, \bibinfo
  {author} {\bibfnamefont {N.-V.-Q.}\ \bibnamefont {Tran}}, \bibinfo {author}
  {\bibfnamefont {A.}~\bibnamefont {{De Rossi}}}, \bibinfo {author}
  {\bibfnamefont {R.}~\bibnamefont {Gabet}}, \ and\ \bibinfo {author}
  {\bibfnamefont {Y.}~\bibnamefont {Jaou\"{e}n}},\ }\href {\doibase
  10.1103/PhysRevLett.102.253903} {\bibfield  {journal} {\bibinfo  {journal}
  {Phys. Rev. Lett.}\ }\textbf {\bibinfo {volume} {102}},\ \bibinfo {pages}
  {253903} (\bibinfo {year} {2009})}\BibitemShut {NoStop}%
\bibitem [{\citenamefont {Fussell}\ \emph {et~al.}(2008)\citenamefont
  {Fussell}, \citenamefont {Hughes},\ and\ \citenamefont
  {Dignam}}]{Fussell2008}%
  \BibitemOpen
  \bibfield  {author} {\bibinfo {author} {\bibfnamefont {D.~P.}\ \bibnamefont
  {Fussell}}, \bibinfo {author} {\bibfnamefont {S.}~\bibnamefont {Hughes}}, \
  and\ \bibinfo {author} {\bibfnamefont {M.~M.}\ \bibnamefont {Dignam}},\
  }\href {\doibase 10.1103/PhysRevB.78.144201} {\bibfield  {journal} {\bibinfo
  {journal} {Phys. Rev. B}\ }\textbf {\bibinfo {volume} {78}},\ \bibinfo
  {pages} {144201} (\bibinfo {year} {2008})}\BibitemShut {NoStop}%
\bibitem [{\citenamefont {{Ba Hoang}}\ \emph {et~al.}(2012)\citenamefont {{Ba
  Hoang}}, \citenamefont {Beetz}, \citenamefont {Midolo}, \citenamefont
  {Skacel}, \citenamefont {Lermer}, \citenamefont {Kamp}, \citenamefont
  {H\"{o}̈fling}, \citenamefont {Balet}, \citenamefont {Chauvin},\ and\
  \citenamefont {Fiore}}]{BaHoang2012}%
  \BibitemOpen
  \bibfield  {author} {\bibinfo {author} {\bibfnamefont {T.}~\bibnamefont {{Ba
  Hoang}}}, \bibinfo {author} {\bibfnamefont {J.}~\bibnamefont {Beetz}},
  \bibinfo {author} {\bibfnamefont {L.}~\bibnamefont {Midolo}}, \bibinfo
  {author} {\bibfnamefont {M.}~\bibnamefont {Skacel}}, \bibinfo {author}
  {\bibfnamefont {M.}~\bibnamefont {Lermer}}, \bibinfo {author} {\bibfnamefont
  {M.}~\bibnamefont {Kamp}}, \bibinfo {author} {\bibfnamefont {S.}~\bibnamefont
  {H\"{o}̈fling}}, \bibinfo {author} {\bibfnamefont {L.}~\bibnamefont
  {Balet}}, \bibinfo {author} {\bibfnamefont {N.}~\bibnamefont {Chauvin}}, \
  and\ \bibinfo {author} {\bibfnamefont {A.}~\bibnamefont {Fiore}},\ }\href
  {\doibase 10.1063/1.3683541} {\bibfield  {journal} {\bibinfo  {journal}
  {Appl. Phys. Lett.}\ }\textbf {\bibinfo {volume} {100}},\ \bibinfo {pages}
  {061122} (\bibinfo {year} {2012})}\BibitemShut {NoStop}%
\bibitem [{\citenamefont {Dubrovskii}\ \emph {et~al.}(2009)\citenamefont
  {Dubrovskii}, \citenamefont {Cirlin},\ and\ \citenamefont
  {Ustinov}}]{Dubrovskii2009}%
  \BibitemOpen
  \bibfield  {author} {\bibinfo {author} {\bibfnamefont {V.~G.}\ \bibnamefont
  {Dubrovskii}}, \bibinfo {author} {\bibfnamefont {G.~E.}\ \bibnamefont
  {Cirlin}}, \ and\ \bibinfo {author} {\bibfnamefont {V.~M.}\ \bibnamefont
  {Ustinov}},\ }\href {\doibase 10.1134/S106378260912001X} {\bibfield
  {journal} {\bibinfo  {journal} {Semiconductors}\ }\textbf {\bibinfo {volume}
  {43}},\ \bibinfo {pages} {1539} (\bibinfo {year} {2009})}\BibitemShut
  {NoStop}%
\bibitem [{\citenamefont {Joyce}\ \emph {et~al.}(2011)\citenamefont {Joyce},
  \citenamefont {Gao}, \citenamefont {{Hoe Tan}}, \citenamefont {Jagadish},
  \citenamefont {Kim}, \citenamefont {Zou}, \citenamefont {Smith},
  \citenamefont {Jackson}, \citenamefont {Yarrison-Rice}, \citenamefont
  {Parkinson},\ and\ \citenamefont {Johnston}}]{Joyce2011}%
  \BibitemOpen
  \bibfield  {author} {\bibinfo {author} {\bibfnamefont {H.~J.}\ \bibnamefont
  {Joyce}}, \bibinfo {author} {\bibfnamefont {Q.}~\bibnamefont {Gao}}, \bibinfo
  {author} {\bibfnamefont {H.}~\bibnamefont {{Hoe Tan}}}, \bibinfo {author}
  {\bibfnamefont {C.}~\bibnamefont {Jagadish}}, \bibinfo {author}
  {\bibfnamefont {Y.}~\bibnamefont {Kim}}, \bibinfo {author} {\bibfnamefont
  {J.}~\bibnamefont {Zou}}, \bibinfo {author} {\bibfnamefont {L.~M.}\
  \bibnamefont {Smith}}, \bibinfo {author} {\bibfnamefont {H.~E.}\ \bibnamefont
  {Jackson}}, \bibinfo {author} {\bibfnamefont {J.~M.}\ \bibnamefont
  {Yarrison-Rice}}, \bibinfo {author} {\bibfnamefont {P.}~\bibnamefont
  {Parkinson}}, \ and\ \bibinfo {author} {\bibfnamefont {M.~B.}\ \bibnamefont
  {Johnston}},\ }\href {\doibase 10.1016/j.pquantelec.2011.03.002} {\bibfield
  {journal} {\bibinfo  {journal} {Progress in Quantum Electronics}\ }\textbf
  {\bibinfo {volume} {35}},\ \bibinfo {pages} {23} (\bibinfo {year}
  {2011})}\BibitemShut {NoStop}%
\bibitem [{\citenamefont {Angelatos}\ and\ \citenamefont
  {Hughes}(2014)}]{Angelatos2014}%
  \BibitemOpen
  \bibfield  {author} {\bibinfo {author} {\bibfnamefont {G.}~\bibnamefont
  {Angelatos}}\ and\ \bibinfo {author} {\bibfnamefont {S.}~\bibnamefont
  {Hughes}},\ }\href {\doibase 10.1103/PhysRevB.90.205406} {\bibfield
  {journal} {\bibinfo  {journal} {Phys. Rev. B}\ }\textbf {\bibinfo {volume}
  {90}},\ \bibinfo {pages} {205406} (\bibinfo {year} {2014})}\BibitemShut
  {NoStop}%
\bibitem [{\citenamefont {Angelatos}\ and\ \citenamefont
  {Hughes}(2015)}]{Angelatos2015}%
  \BibitemOpen
  \bibfield  {author} {\bibinfo {author} {\bibfnamefont {G.}~\bibnamefont
  {Angelatos}}\ and\ \bibinfo {author} {\bibfnamefont {S.}~\bibnamefont
  {Hughes}},\ }\href {\doibase 10.1103/PhysRevA.91.051803} {\bibfield
  {journal} {\bibinfo  {journal} {Phys. Rev. A}\ }\textbf {\bibinfo {volume}
  {91}},\ \bibinfo {pages} {051803(R)} (\bibinfo {year} {2015})}\BibitemShut
  {NoStop}%
\bibitem [{\citenamefont {Makhonin}\ \emph {et~al.}(2013)\citenamefont
  {Makhonin}, \citenamefont {Foster}, \citenamefont {Krysa}, \citenamefont
  {Fry}, \citenamefont {Davies}, \citenamefont {Grange}, \citenamefont
  {Walther}, \citenamefont {Skolnick},\ and\ \citenamefont
  {Wilson}}]{Makhonin2013}%
  \BibitemOpen
  \bibfield  {author} {\bibinfo {author} {\bibfnamefont {M.~N.}\ \bibnamefont
  {Makhonin}}, \bibinfo {author} {\bibfnamefont {A.~P.}\ \bibnamefont
  {Foster}}, \bibinfo {author} {\bibfnamefont {A.~B.}\ \bibnamefont {Krysa}},
  \bibinfo {author} {\bibfnamefont {P.~W.}\ \bibnamefont {Fry}}, \bibinfo
  {author} {\bibfnamefont {D.~G.}\ \bibnamefont {Davies}}, \bibinfo {author}
  {\bibfnamefont {T.}~\bibnamefont {Grange}}, \bibinfo {author} {\bibfnamefont
  {T.}~\bibnamefont {Walther}}, \bibinfo {author} {\bibfnamefont {M.~S.}\
  \bibnamefont {Skolnick}}, \ and\ \bibinfo {author} {\bibfnamefont {L.~R.}\
  \bibnamefont {Wilson}},\ }\href {\doibase 10.1021/nl303075q} {\bibfield
  {journal} {\bibinfo  {journal} {Nano Lett.}\ }\textbf {\bibinfo {volume}
  {13}},\ \bibinfo {pages} {861} (\bibinfo {year} {2013})}\BibitemShut
  {NoStop}%
\bibitem [{\citenamefont {Diedenhofen}\ \emph {et~al.}(2011)\citenamefont
  {Diedenhofen}, \citenamefont {Janssen}, \citenamefont {Hocevar},
  \citenamefont {Pierret}, \citenamefont {Bakkers}, \citenamefont {Urbach},\
  and\ \citenamefont {{G\'{o}mez Rivas}}}]{Diedenhofen2011}%
  \BibitemOpen
  \bibfield  {author} {\bibinfo {author} {\bibfnamefont {S.~L.}\ \bibnamefont
  {Diedenhofen}}, \bibinfo {author} {\bibfnamefont {O.~T.~A.}\ \bibnamefont
  {Janssen}}, \bibinfo {author} {\bibfnamefont {M.}~\bibnamefont {Hocevar}},
  \bibinfo {author} {\bibfnamefont {A.}~\bibnamefont {Pierret}}, \bibinfo
  {author} {\bibfnamefont {E.~P. A.~M.}\ \bibnamefont {Bakkers}}, \bibinfo
  {author} {\bibfnamefont {H.~P.}\ \bibnamefont {Urbach}}, \ and\ \bibinfo
  {author} {\bibfnamefont {J.}~\bibnamefont {{G\'{o}mez Rivas}}},\ }\href
  {\doibase 10.1021/nn201557h} {\bibfield  {journal} {\bibinfo  {journal} {ACS
  Nano}\ }\textbf {\bibinfo {volume} {5}},\ \bibinfo {pages} {5830} (\bibinfo
  {year} {2011})}\BibitemShut {NoStop}%
\bibitem [{\citenamefont {Citrin}(2004)}]{Citrin2004}%
  \BibitemOpen
  \bibfield  {author} {\bibinfo {author} {\bibfnamefont {D.~S.}\ \bibnamefont
  {Citrin}},\ }\href {http://pubs.acs.org/doi/abs/10.1021/nl049679l} {\bibfield
   {journal} {\bibinfo  {journal} {Nano Lett.}\ }\textbf {\bibinfo {volume}
  {4}} (\bibinfo {year} {2004})}\BibitemShut {NoStop}%
\bibitem [{\citenamefont {Li}\ \emph {et~al.}(2009)\citenamefont {Li},
  \citenamefont {Evers},\ and\ \citenamefont {Keitel}}]{Li2009}%
  \BibitemOpen
  \bibfield  {author} {\bibinfo {author} {\bibfnamefont {G.~X.}\ \bibnamefont
  {Li}}, \bibinfo {author} {\bibfnamefont {J.}~\bibnamefont {Evers}}, \ and\
  \bibinfo {author} {\bibfnamefont {C.~H.}\ \bibnamefont {Keitel}},\ }\href
  {\doibase 10.1103/PhysRevB.80.045102} {\bibfield  {journal} {\bibinfo
  {journal} {Phys. Rev. B - Condens. Matter Mater. Phys.}\ }\textbf {\bibinfo
  {volume} {80}},\ \bibinfo {pages} {1} (\bibinfo {year} {2009})},\ \Eprint
  {http://arxiv.org/abs/0902.2890} {arXiv:0902.2890} \BibitemShut {NoStop}%
\bibitem [{\citenamefont {Yao}\ \emph {et~al.}(2009{\natexlab{b}})\citenamefont
  {Yao}, \citenamefont {{Van Vlack}}, \citenamefont {Reza}, \citenamefont
  {Patterson}, \citenamefont {Dignam},\ and\ \citenamefont
  {Hughes}}]{Yao2009meta}%
  \BibitemOpen
  \bibfield  {author} {\bibinfo {author} {\bibfnamefont {P.}~\bibnamefont
  {Yao}}, \bibinfo {author} {\bibfnamefont {C.~P.}\ \bibnamefont {{Van
  Vlack}}}, \bibinfo {author} {\bibfnamefont {A.}~\bibnamefont {Reza}},
  \bibinfo {author} {\bibfnamefont {M.}~\bibnamefont {Patterson}}, \bibinfo
  {author} {\bibfnamefont {M.~M.}\ \bibnamefont {Dignam}}, \ and\ \bibinfo
  {author} {\bibfnamefont {S.}~\bibnamefont {Hughes}},\ }\href {\doibase
  10.1103/PhysRevB.80.195106} {\bibfield  {journal} {\bibinfo  {journal} {Phys.
  Rev. B}\ }\textbf {\bibinfo {volume} {80}},\ \bibinfo {pages} {195106}
  (\bibinfo {year} {2009}{\natexlab{b}})}\BibitemShut {NoStop}%
\bibitem [{\citenamefont {Christ}\ \emph {et~al.}(2003)\citenamefont {Christ},
  \citenamefont {Tikhodeev}, \citenamefont {Gippius},\ and\ \citenamefont
  {Giessen}}]{Christ2003}%
  \BibitemOpen
  \bibfield  {author} {\bibinfo {author} {\bibfnamefont {A.}~\bibnamefont
  {Christ}}, \bibinfo {author} {\bibfnamefont {S.~G.}\ \bibnamefont
  {Tikhodeev}}, \bibinfo {author} {\bibfnamefont {J.}~\bibnamefont {Gippius},
  \bibfnamefont {N.~A .and~Kuhl}}, \ and\ \bibinfo {author} {\bibfnamefont
  {H.}~\bibnamefont {Giessen}},\ }\href {\doibase
  10.1103/PhysRevLett.91.183901} {\bibfield  {journal} {\bibinfo  {journal}
  {Phys. Rev. Lett.}\ }\textbf {\bibinfo {volume} {91}},\ \bibinfo {pages}
  {183901} (\bibinfo {year} {2003})}\BibitemShut {NoStop}%
\bibitem [{\citenamefont {F\'{e}vrier}\ \emph {et~al.}(2012)\citenamefont
  {F\'{e}vrier}, \citenamefont {Gogol}, \citenamefont {Aassime}, \citenamefont
  {M\'{e}gy}, \citenamefont {Delacour}, \citenamefont {Chelnokov},
  \citenamefont {Apuzzo}, \citenamefont {Blaize}, \citenamefont {Lourtioz},\
  and\ \citenamefont {Dagens}}]{Fevrier2012}%
  \BibitemOpen
  \bibfield  {author} {\bibinfo {author} {\bibfnamefont {M.}~\bibnamefont
  {F\'{e}vrier}}, \bibinfo {author} {\bibfnamefont {P.}~\bibnamefont {Gogol}},
  \bibinfo {author} {\bibfnamefont {A.}~\bibnamefont {Aassime}}, \bibinfo
  {author} {\bibfnamefont {R.}~\bibnamefont {M\'{e}gy}}, \bibinfo {author}
  {\bibfnamefont {C.}~\bibnamefont {Delacour}}, \bibinfo {author}
  {\bibfnamefont {A.}~\bibnamefont {Chelnokov}}, \bibinfo {author}
  {\bibfnamefont {A.}~\bibnamefont {Apuzzo}}, \bibinfo {author} {\bibfnamefont
  {S.}~\bibnamefont {Blaize}}, \bibinfo {author} {\bibfnamefont {J.~M.}\
  \bibnamefont {Lourtioz}}, \ and\ \bibinfo {author} {\bibfnamefont
  {B.}~\bibnamefont {Dagens}},\ }\href {\doibase 10.1021/nl204265f} {\bibfield
  {journal} {\bibinfo  {journal} {Nano Lett.}\ }\textbf {\bibinfo {volume}
  {12}},\ \bibinfo {pages} {1032} (\bibinfo {year} {2012})}\BibitemShut
  {NoStop}%
\bibitem [{\citenamefont {Yablonskii}\ \emph {et~al.}(2001)\citenamefont
  {Yablonskii}, \citenamefont {Muljarov}, \citenamefont {Gippius},
  \citenamefont {Tikhodeev}, \citenamefont {Fujita},\ and\ \citenamefont
  {Ishihara}}]{Yablonskii2001}%
  \BibitemOpen
  \bibfield  {author} {\bibinfo {author} {\bibfnamefont {A.~L.}\ \bibnamefont
  {Yablonskii}}, \bibinfo {author} {\bibfnamefont {E.~A.}\ \bibnamefont
  {Muljarov}}, \bibinfo {author} {\bibfnamefont {N.~A.}\ \bibnamefont
  {Gippius}}, \bibinfo {author} {\bibfnamefont {S.~G.}\ \bibnamefont
  {Tikhodeev}}, \bibinfo {author} {\bibfnamefont {T.}~\bibnamefont {Fujita}}, \
  and\ \bibinfo {author} {\bibfnamefont {T.}~\bibnamefont {Ishihara}},\ }\href
  {\doibase 10.1143/JPSJ.70.1137} {\bibfield  {journal} {\bibinfo  {journal}
  {J. Phys. Soc. Japan}\ }\textbf {\bibinfo {volume} {70}},\ \bibinfo {pages}
  {1137} (\bibinfo {year} {2001})}\BibitemShut {NoStop}%
\bibitem [{\citenamefont {Tokushima}\ \emph {et~al.}(2004)\citenamefont
  {Tokushima}, \citenamefont {Yamada},\ and\ \citenamefont
  {Arakawa}}]{Tokushima2004}%
  \BibitemOpen
  \bibfield  {author} {\bibinfo {author} {\bibfnamefont {M.}~\bibnamefont
  {Tokushima}}, \bibinfo {author} {\bibfnamefont {H.}~\bibnamefont {Yamada}}, \
  and\ \bibinfo {author} {\bibfnamefont {Y.}~\bibnamefont {Arakawa}},\ }\href
  {\doibase 10.1063/1.1755838} {\bibfield  {journal} {\bibinfo  {journal}
  {Appl. Phys. Lett.}\ }\textbf {\bibinfo {volume} {84}},\ \bibinfo {pages}
  {4298} (\bibinfo {year} {2004})}\BibitemShut {NoStop}%
\bibitem [{\citenamefont {Weiler}\ \emph {et~al.}(2012)\citenamefont {Weiler},
  \citenamefont {Ulhaq}, \citenamefont {Ulrich}, \citenamefont {Richter},
  \citenamefont {Jetter}, \citenamefont {Michler}, \citenamefont {Roy},\ and\
  \citenamefont {Hughes}}]{Weiler2012}%
  \BibitemOpen
  \bibfield  {author} {\bibinfo {author} {\bibfnamefont {S.}~\bibnamefont
  {Weiler}}, \bibinfo {author} {\bibfnamefont {A.}~\bibnamefont {Ulhaq}},
  \bibinfo {author} {\bibfnamefont {S.~M.}\ \bibnamefont {Ulrich}}, \bibinfo
  {author} {\bibfnamefont {D.}~\bibnamefont {Richter}}, \bibinfo {author}
  {\bibfnamefont {M.}~\bibnamefont {Jetter}}, \bibinfo {author} {\bibfnamefont
  {P.}~\bibnamefont {Michler}}, \bibinfo {author} {\bibfnamefont
  {C.}~\bibnamefont {Roy}}, \ and\ \bibinfo {author} {\bibfnamefont
  {S.}~\bibnamefont {Hughes}},\ }\href {\doibase 10.1103/PhysRevB.86.241304}
  {\bibfield  {journal} {\bibinfo  {journal} {Phys. Rev. B}\ }\textbf {\bibinfo
  {volume} {86}},\ \bibinfo {pages} {241304} (\bibinfo {year}
  {2012})}\BibitemShut {NoStop}%
\bibitem [{\citenamefont {Novotny}\ and\ \citenamefont
  {Hecht}(2006)}]{Novotny2006}%
  \BibitemOpen
  \bibfield  {author} {\bibinfo {author} {\bibfnamefont {L.}~\bibnamefont
  {Novotny}}\ and\ \bibinfo {author} {\bibfnamefont {B.}~\bibnamefont
  {Hecht}},\ }\href
  {http://books.google.ca/books/about/Principles\_of\_Nano\_Optics.html?id=Qrf036kThTQC\&pgis=1}
  {\emph {\bibinfo {title} {{Principles of Nano-Optics}}}}\ (\bibinfo
  {publisher} {Cambridge University Press},\ \bibinfo {address} {Cambridge},\
  \bibinfo {year} {2006})\BibitemShut {NoStop}%
\bibitem [{\citenamefont {Mahmoodian}\ \emph {et~al.}(2009)\citenamefont
  {Mahmoodian}, \citenamefont {Poulton}, \citenamefont {Dossou}, \citenamefont
  {McPhedran}, \citenamefont {Botten},\ and\ \citenamefont
  {de~Sterke}}]{Mahmoodian2009}%
  \BibitemOpen
  \bibfield  {author} {\bibinfo {author} {\bibfnamefont {S.}~\bibnamefont
  {Mahmoodian}}, \bibinfo {author} {\bibfnamefont {C.~G.}\ \bibnamefont
  {Poulton}}, \bibinfo {author} {\bibfnamefont {K.~B.}\ \bibnamefont {Dossou}},
  \bibinfo {author} {\bibfnamefont {R.~C.}\ \bibnamefont {McPhedran}}, \bibinfo
  {author} {\bibfnamefont {L.~C.}\ \bibnamefont {Botten}}, \ and\ \bibinfo
  {author} {\bibfnamefont {C.~M.}\ \bibnamefont {de~Sterke}},\ }\href {\doibase
  10.1364/OE.17.019629} {\bibfield  {journal} {\bibinfo  {journal} {Opt.
  Express}\ }\textbf {\bibinfo {volume} {17}},\ \bibinfo {pages} {19629}
  (\bibinfo {year} {2009})}\BibitemShut {NoStop}%
\bibitem [{\citenamefont {Patterson}\ and\ \citenamefont
  {Hughes}(2010)}]{Patterson2010}%
  \BibitemOpen
  \bibfield  {author} {\bibinfo {author} {\bibfnamefont {M.}~\bibnamefont
  {Patterson}}\ and\ \bibinfo {author} {\bibfnamefont {S.}~\bibnamefont
  {Hughes}},\ }\href {\doibase 10.1103/PhysRevB.81.245321} {\bibfield
  {journal} {\bibinfo  {journal} {Phys. Rev. B}\ }\textbf {\bibinfo {volume}
  {81}},\ \bibinfo {pages} {245321} (\bibinfo {year} {2010})}\BibitemShut
  {NoStop}%
\bibitem [{\citenamefont {Angelatos}(2015)}]{Angelatos2015th}%
  \BibitemOpen
  \bibfield  {author} {\bibinfo {author} {\bibfnamefont {G.}~\bibnamefont
  {Angelatos}},\ }\emph {\bibinfo {title} {Theory and Applications of
  Light-matter interactions in quantum dot nanowire photonic crystal
  systems}},\ \href@noop {} {\bibinfo {type} {Master's thesis}},\ \bibinfo
  {school} {Queen's University} (\bibinfo {year} {2015})\BibitemShut {NoStop}%
\bibitem [{\citenamefont {{Manga Rao}}\ and\ \citenamefont
  {Hughes}(2007)}]{Rao2007theory}%
  \BibitemOpen
  \bibfield  {author} {\bibinfo {author} {\bibfnamefont {V.~S.~C.}\
  \bibnamefont {{Manga Rao}}}\ and\ \bibinfo {author} {\bibfnamefont
  {S.}~\bibnamefont {Hughes}},\ }\href {\doibase 10.1103/PhysRevB.75.205437}
  {\bibfield  {journal} {\bibinfo  {journal} {Phys. Rev. B}\ }\textbf {\bibinfo
  {volume} {75}},\ \bibinfo {pages} {205437} (\bibinfo {year}
  {2007})}\BibitemShut {NoStop}%
\bibitem [{\citenamefont {Kristensen}\ \emph {et~al.}(2011)\citenamefont
  {Kristensen}, \citenamefont {M\o{}rk}, \citenamefont {Lodahl},\ and\
  \citenamefont {Hughes}}]{Kristensen2011}%
  \BibitemOpen
  \bibfield  {author} {\bibinfo {author} {\bibfnamefont {P.~T.}\ \bibnamefont
  {Kristensen}}, \bibinfo {author} {\bibfnamefont {J.}~\bibnamefont {M\o{}rk}},
  \bibinfo {author} {\bibfnamefont {P.}~\bibnamefont {Lodahl}}, \ and\ \bibinfo
  {author} {\bibfnamefont {S.}~\bibnamefont {Hughes}},\ }\href {\doibase
  10.1103/PhysRevB.83.075305} {\bibfield  {journal} {\bibinfo  {journal} {Phys.
  Rev. B}\ }\textbf {\bibinfo {volume} {83}},\ \bibinfo {pages} {075305}
  (\bibinfo {year} {2011})}\BibitemShut {NoStop}%
\bibitem [{\citenamefont {Suttorp}\ and\ \citenamefont
  {Wonderen}(2004)}]{Suttorp2004}%
  \BibitemOpen
  \bibfield  {author} {\bibinfo {author} {\bibfnamefont {L.~G.}\ \bibnamefont
  {Suttorp}}\ and\ \bibinfo {author} {\bibfnamefont {A.~J.~V.}\ \bibnamefont
  {Wonderen}},\ }\href {\doibase 10.1209/epl/i2004-10131-8} {\bibfield
  {journal} {\bibinfo  {journal} {Europhys. Lett.}\ }\textbf {\bibinfo {volume}
  {67}},\ \bibinfo {pages} {766} (\bibinfo {year} {2004})}\BibitemShut
  {NoStop}%
\bibitem [{\citenamefont {Meystre}\ and\ \citenamefont
  {Sargent}(1999)}]{Meystre1999}%
  \BibitemOpen
  \bibfield  {author} {\bibinfo {author} {\bibfnamefont {P.}~\bibnamefont
  {Meystre}}\ and\ \bibinfo {author} {\bibfnamefont {M.}~\bibnamefont
  {Sargent}},\ }\href
  {http://books.google.ca/books/about/Elements\_of\_Quantum\_Optics.html?id=dWnIOHloxoEC\&pgis=1}
  {\emph {\bibinfo {title} {{Elements of Quantum Optics}}}}\ (\bibinfo
  {publisher} {Springer Science \& Business Media},\ \bibinfo {year} {1999})\
  p.\ \bibinfo {pages} {432}\BibitemShut {NoStop}%
\bibitem [{\citenamefont {Fussell}\ and\ \citenamefont
  {Dignam}(2007)}]{Fussell2007}%
  \BibitemOpen
  \bibfield  {author} {\bibinfo {author} {\bibfnamefont {D.~P.}\ \bibnamefont
  {Fussell}}\ and\ \bibinfo {author} {\bibfnamefont {M.~M.}\ \bibnamefont
  {Dignam}},\ }\href {\doibase 10.1103/PhysRevA.76.053801} {\bibfield
  {journal} {\bibinfo  {journal} {Phys. Rev. A}\ }\textbf {\bibinfo {volume}
  {76}},\ \bibinfo {pages} {053801} (\bibinfo {year} {2007})}\BibitemShut
  {NoStop}%
\end{thebibliography}%
\end{document}